\documentclass[reprint]{revtex4-1}
\usepackage[utf8]{inputenc}
\usepackage{hyperref}
\usepackage{graphicx}
\usepackage{physics}
\usepackage{float}
\usepackage{subfiles}

\usepackage[dvipsnames]{xcolor}

\newcommand*{\balancecolsandclearpage}{%
  \close@column@grid
  \cleardoublepage
  \twocolumngrid
}

\begin{document}

\title{Analyzing the Performance of Variational Quantum Factoring on a Superconducting Quantum Processor}

\date{\today}

\author{Amir H. Karamlou$^{1,2,}$}
\thanks{These authors contributed equally.}

\author{William A. Simon$^{1,}$}
\thanks{These authors contributed equally.}

\author{Amara Katabarwa$^1$}
\author{Travis L. Scholten$^3$}
\author{Borja Peropadre$^1$}
\author{Yudong Cao$^{1,}$}
\email{yudong@zapatacomputing.com}

\affiliation{$^1$Zapata Computing, Boston, MA 02110 USA}

\affiliation{$^2$Research Laboratory of Electronics, Massachusetts Institute of Technology, Cambridge, MA 02139, USA}

\affiliation{$^3$IBM Quantum, IBM T. J. Watson Research Center, Yorktown Heights, NY 10598}

\begin{abstract}

    In the near-term, hybrid quantum-classical algorithms hold great potential for outperforming classical approaches. Understanding how these two computing paradigms work in tandem is critical for identifying areas where such hybrid algorithms could provide a quantum advantage. In this work, we study a QAOA-based quantum optimization algorithm by implementing the Variational Quantum Factoring (VQF) algorithm. We execute experimental demonstrations using a superconducting quantum processor, and investigate the trade off between quantum resources (number of qubits and circuit depth) and the probability that a given biprime is successfully factored. In our experiments, the integers 1099551473989, 3127, and 6557 are factored with 3, 4, and 5 qubits, respectively, using a QAOA ansatz with up to 8 layers and we are able to identify the optimal number of circuit layers for a given instance to maximize success probability. Furthermore, we demonstrate the impact of different noise sources on the performance of QAOA, and reveal the coherent error caused by the residual $ZZ$-coupling between qubits as a dominant source of error in the superconducting quantum processor.

\end{abstract}


\maketitle

\section{Introduction}

\begin{figure*}[t]
    \centering
    \includegraphics[width=17.5cm]{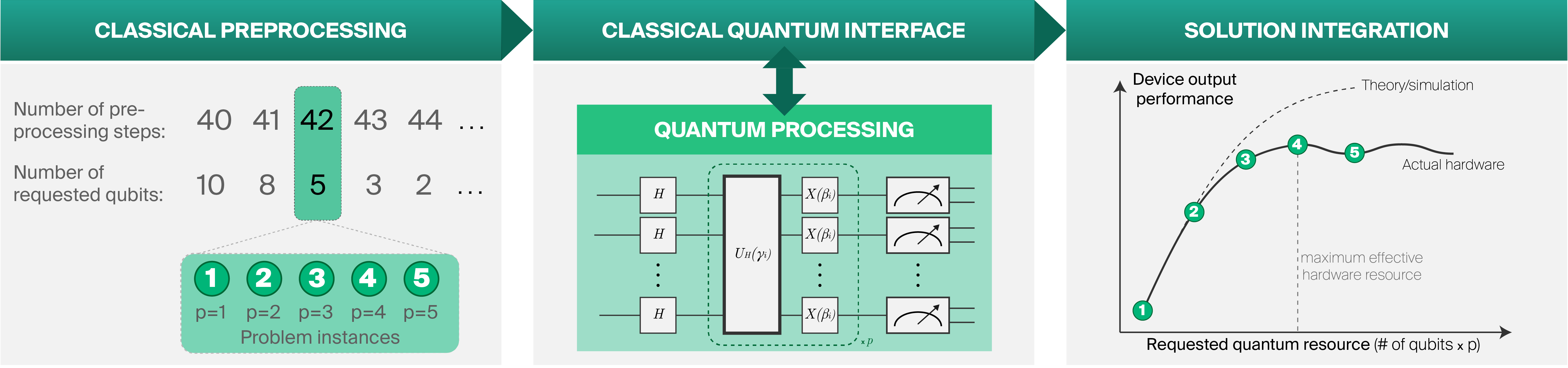}
    \caption{\textbf{Tunable resource tradeoffs with variational quantum factoring (VQF)}. \underline{Left:} Given an integer $N$ to be factored, varying amounts of classical pre-processing steps result in a different number of qubits required for the optimization problem, and defines a ``problem Hamiltonian". \underline{Middle:} The optimization problem is solved on quantum hardware using the QAOA with $p$ layers. Using classical optimization, the algorithm finds parameters $\boldsymbol{\gamma}$ and $\boldsymbol{\beta}$ that prepare a trial state which approximates the ground state of the problem Hamiltonian. \underline{Right}: Classical post-processing combines the measurement results on the quantum device with classical pre-processing and evaluates the algorithm success.}
    \label{fig:concept}
\end{figure*}

While near-term quantum devices are approaching the limits of classical tractability \cite{Arute2019, Zhong2020}, they are limited in the number of physical qubits, and can only execute finite-depth circuits with sufficient fidelity to be useful in applications \cite{Kjaergaard}. Hybrid quantum-classical algorithms hold great promise for achieving a meaningful quantum advantage in the near-term \cite{Google2020, Peruzzo2014, havlicek2019} by combining the quantum resources offered by near-term devices with the computational power of classical processors. In a hybrid algorithm, a classical computer pre-processes a problem instance to expose the core components which capture its essential computational hardness, and is also in a form compatible with a quantum algorithm. The classical computer then utilizes a near-term quantum computer to finish solving the problem. The output of the quantum computer is classical bitstrings which are post-processed classically. This interaction may be iterative: the classical computer may also send communication and control signals to and from the quantum device, for example, by proposing new parameter values to be used in a parameterized quantum circuit.

Quantum algorithms developed for near-term quantum hardware will have to contend with several hardware restrictions, such as number of qubits, coupling connectivity, gate fidelities, and various sources of noise, which are all highly dependent on the particular quantum processor the algorithm is being run on. Tradeoffs between different resources have recently been demonstrated in the context of running quantum circuits \cite{Proctor2020}. However, even as fault-tolerant quantum processors are built, classical computing will still be required to perform pre-processing and post-processing, as well as for quantum error-correction \cite{gottesman2009}. Therefore, understanding how quantum and classical computing resources can be leveraged together is imperative for extracting maximal utility from quantum computers.

An illustration of this hybrid scheme, in the context of this work, is shown in Fig. \ref{fig:concept}. In this work, we implement the \emph{variational quantum factoring} (VQF) algorithm \cite{Anschuetz2018} on a superconducting quantum processor \cite{Krantz}. VQF is an algorithm for tackling an NP problem with a tunable trade off between classical and quantum computing resources, and is feasible for near-term devices. VQF uses classical pre-processing to reduce integer factoring to a combinatorial optimization problem which can be solved using the  quantum approximate optimization algorithm (QAOA) \cite{Farhi2014}.

VQF is a useful algorithm to study for several reasons. First, an inverse relationship exists between the amount of classical pre-processing and the number of required qubits needed to finish solving the problem, resulting in a tunable tradeoff between the amount of classical computation and qubits used. Second, the complexity of the quantum circuit acting on those qubits is tunable by adjusting the number of layers used in the QAOA ansatz. Finally, VQF has the advantage that its success is quickly verifiable on a classical computer, by checking if the proposed factors multiply to the input biprime. 

This paper is organized as follows. Section \ref{sec:VQF} provides an overview of the variational quantum factoring algorithm. Section \ref{sec:experiment} describes the experimental results from our implementation. Furthermore, we study the impact of different sources of noise on the algorithm's performance. The analysis in this section extends beyond VQF and to any general instance of QAOA.  We conclude in Section \ref{sec:conclusion}, and consider the implications of this work.

\section{Variational Quantum Factoring}
\label{sec:VQF}
The variational quantum factoring (VQF) algorithm maps the problem of factoring into an optimization problem \cite{Burges2002}. Given an $n$-bit biprime number \begin{equation}
    N=\sum_{k=0}^{n-1} 2^k N_k,
\end{equation} factoring involves finding the two prime factors $p$ and $q$ satisfying $N = pq$, where
\begin{align}
    p&=\sum_{k=0}^{n_p-1} 2^k p_k, \\
    q&=\sum_{k=0}^{n_q-1} 2^k q_k.
\end{align}
That is, $p$ and $q$ can be represented with $n_q$ and $n_p$ bits, respectively.

Factoring in this way can be thought of as the inverse problem to the ``longhand" binary multiplication:  the value of the $i^{\mathrm{th}}$ bit of the result, $N_i$, is known and the task is to solve for the bits of the prime factors $\{p_i\}$ and $\{q_i\}$. An explicit binary multiplication of $p$ and $q$ yields a series of equations that have to be satisfied by $\{p_i\}$ and $\{q_i\}$, along with carry bits $\{z_{i,j}\}$ which denote a bit carry from the $i^{\mathrm{th}}$ to the $j^{\mathrm{th}}$ position. Re-writing each equation in the series allows it to be associated to a particular \emph{clause} in an optimization problem relating to the bit $N_{i}$:
\begin{equation}\label{eq:clauses}
    C_i=N_i - \sum_{j=0}^i q_j p_{i-j} - \sum_{j=0}^i z_{i,j} + \sum_{j=1}^{n_p+n_q-1} 2^j z_{i,i+j}.
\end{equation}
In order for each clause, $C_{i}$, to be 0, the values of $\{p_i\}$, $\{q_i\}$ and the carry bits $\{z_{i,j}\}$ in the clause must be correct. By satisfying the constraint for all clauses, the factors $p$ and $q$ can be retrieved. 

Given this, the problem of factoring is thus reduced to a combinatorial  optimization problem. Such problems can be solved using quantum computers by associating a qubit with each bit in the cost function. The number of qubits needed depends on the number of bits in the clauses. As discussed in \cite{Anschuetz2018}, some number of \emph{classical} preprocessing \emph{heuristics} can be used to simplify the clauses (that is, assign values to some of the bits $\{p_{i}\}$ and $\{q_{i}\}$). As classical preprocessing removes variables from the optimization problem (by explicitly assigning bit values), the number of qubits needed to complete the solution to the problem is reduced.
The new set of clauses will be denoted as $\{C^\prime_i\}$; Appendix \ref{appendix:preprocessing} discusses details of this classical preprocessing.

Each clause $C^\prime_i$ can be mapped to a term in an Ising Hamiltonian $\hat{C}_i$ by associating each bit value ($b_k \in \{p_{i}, q_{i}, z_{i,j}\}$) to a corresponding qubit operator:
\begin{equation}
    b_k \mapsto \frac{1}{2}(1-Z_k).
\end{equation}
The solution to the factoring problem corresponds to finding the ground state of the Hamiltonian
\begin{equation} \label{eq:Hamiltonian}
    \hat{H}_{C^\prime}=\sum_{i=0}^n \hat{C}_i^2,
\end{equation}
with a well defined ground state energy $E_0=0$. Because each clause $C_i$ in equation \ref{eq:clauses} contains quadratic terms in the bits and the Hamiltonian $\hat{H}_{C^\prime}$ is a sum of squares of $\hat{C}_i$, $\hat{H}_{C^\prime}$ includes 4-local terms of the form $Z_i\otimes Z_j\otimes Z_k\otimes Z_l$. (An operator is $k$-local if it acts non-trivially on at most $k$ qubits.) Much of the literature on quantum optimization so far has considered solving {\sf MAXCUT} problems on $d$-regular graphs \cite{Farhi2014,Guerreschi2019}, which can be mapped to problems of finding the ground states of 2-local Hamiltonians. Other problems, such as {\sf MAX-3-LIN-2} \cite{1412.6062,1905.07047}, can be directly mapped to ground state problems of 3-local Hamiltonians. The 4-local Ising Hamiltonian problems produced for VQF motivates a new entry to the classes of Ising Hamiltonian problems currently studied in the context of QAOA. In principle, one can reduce any $k$-local Hamiltonian to $2$-local by well-known techniques \cite{Biamonte2008,1901.04405}. However, the interactions between the qubits in the resulting 2-local Hamiltonian is not guaranteed to correspond to a $d$-regular graph, which again falls outside the scope of existing considerations in the literature.

The ground state of the Hamiltonian in equation (\ref{eq:Hamiltonian}) can be approximated on near-term, digital quantum computers using QAOA  \cite{Farhi2014}. QAOA is a 
good candidate for demonstrating quantum advantage through combinatorial optimization \cite{Arute2020, Bengtsson2020, Lacroix2020}. Each layer of a QAOA ansatz consists of two unitary operators, each with a tunable parameter. The first unitary operator is
\begin{equation}
    U_H(\gamma)=e^{-i \gamma \hat{H}_{C^\prime}}~\text{where}~\gamma \in [0, 2\pi),
\end{equation}
which applies an entangling phase according to the cost Hamiltonian $\hat{H}_{C^\prime}$. The second unitary is the admixing operation
\begin{equation}
    U_a(\beta)=\prod_i^n e^{-i \beta X_i}~\text{where}~\beta \in [0, \pi),
\end{equation}
which applies a single-qubit rotation around the $X$-axis with angle $2\beta$ to each qubit.

For a given number of layers $p$, the combination of $U_H(\gamma)$ and $U_a(\beta)$ are repeated sequentially with different parameters, generating the ansatz state 
\begin{equation}\label{eq:ansatz}
    \ket{\boldsymbol{\gamma}, \boldsymbol{\beta}} = U_a(\beta_{p-1})U_H(\gamma_{p-1}) \cdots U_a(\beta_{0})U_H(\gamma_{0}) \ket{+}^{\otimes n},
\end{equation}
parameterized by $\boldsymbol{\gamma}=(\gamma_0,\cdots,\gamma_{p-1})$ and $\boldsymbol{\beta}=(\beta_0,\cdots,\beta_{p-1})$. The approximate ground state for the cost Hamiltonian can reached by tuning these $2p$ parameters. Given this ansatz state, a classical optimizer is used to find the optimal parameters $\boldsymbol{\gamma}_{\rm opt}$ and $\boldsymbol{\beta}_{\rm opt}$ that minimize the expected value of the cost Hamiltonian
\begin{equation}\label{eq:cost}
    E(\boldsymbol{\gamma}, \boldsymbol{\beta})=\bra{\boldsymbol{\gamma}, \boldsymbol{\beta}}\hat{H}_{C^\prime}\ket{\boldsymbol{\gamma}, \boldsymbol{\beta}},
\end{equation}
where for each $\boldsymbol{\gamma}$ and $\boldsymbol{\beta}$ the value $E(\boldsymbol{\gamma}, \boldsymbol{\beta})$ is estimated on a quantum computer.

The circuit $\ket{\boldsymbol{\gamma}_{\rm opt}, \boldsymbol{\beta}_{\rm opt}}$ is then prepared on the quantum computer and measured. If the outcome of the measurement satisfies all the clauses, it can be mapped to the remaining unsolved binary variables in $\{p_i\}$, $\{q_i\}$, and $\{z_{i,j}\}$, resulting in the prime factors $p$ and $q$.

The success of VQF is measured by the probability that a measured bitstring encodes the correct factors. We define the set $M_s =\{m_j \}$ consisting of all bitstrings sampled from the quantum computer that satisfy all the clauses in the Hamiltonian, $\hat{H}_{C^\prime}$. We can therefore define the success rate, $s(\boldsymbol{\gamma}, \boldsymbol{\beta})$, as the proportion of the bitstrings sampled from $\ket{\boldsymbol{\gamma}, \boldsymbol{\beta}}$ that satisfy all clauses in $\{C^\prime_i\}$:
\begin{equation}\label{eq:success rate}
    s(\boldsymbol{\gamma}, \boldsymbol{\beta})= \frac{\abs{M_s}}{\abs{M}}, \hspace{2mm} M_s  = \{ m_j \in M | \sum{C_i} = 0 \} 
\end{equation}
where $\abs{M_s}$ is the number of samples satisfying all the clauses and $\abs{M}$ is the total number of measurements sampled.

In order to successfully factor the input biprime, only one such satisfactory bitstring needs to be observed. However, if that bitstring occurs very rarely, then many repeated preparations and measurements of the trial state are necessary. Therefore, a higher value of $s(\boldsymbol{\gamma}, \boldsymbol{\beta})$ is generally preferred. $E(\boldsymbol{\gamma}, \boldsymbol{\beta}) \geq 0$, with the equality condition satisfied if $s(\boldsymbol{\gamma}, \boldsymbol{\beta})=1$.


\begin{figure}[b]
    \centering
    \includegraphics[width=8.5cm]{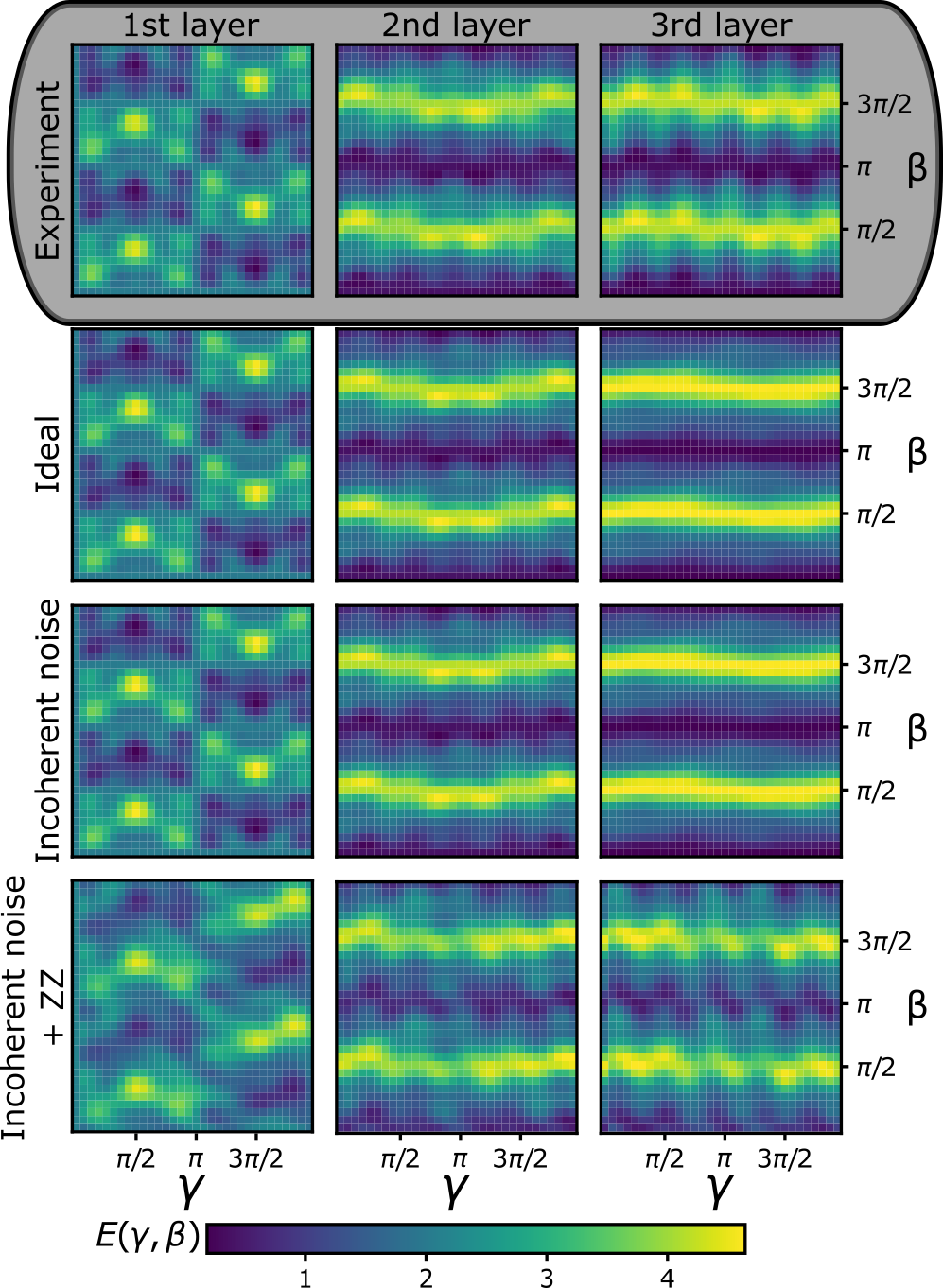}
    \caption{\textbf{QAOA optimization landscapes}. For a fixed VQF instance on 3 qubits, we study the impact of noise (rows) on the optimization landscape as a function of the number of QAOA layers in the ansatz (columns). To produce these landscapes, we use a resolution of $\pi/32$, resulting in 1024 grid points. For the second and third layers, the ansatz parameters for the previous layers are fixed to the optimal parameters found through ideal simulation. } 
    \label{fig:energy landscapes - zz}
\end{figure}

\begin{figure*}[t]
    \centering
    \includegraphics[height=2in]{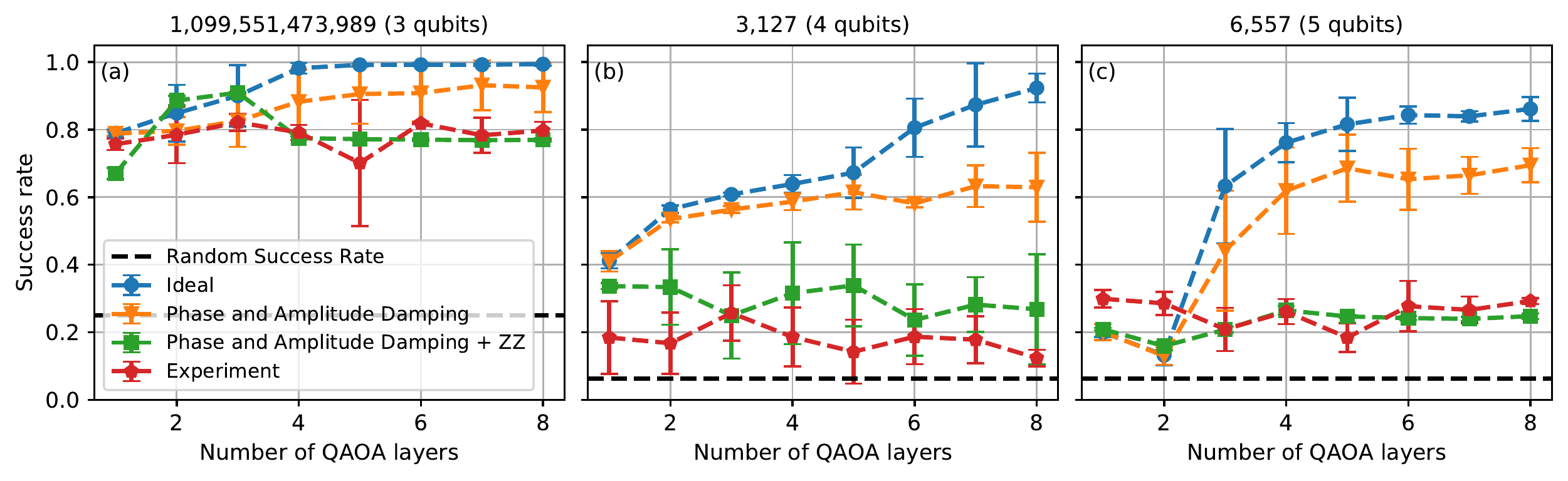}
    \caption{\textbf{Success rates for running VQF on the integers (a) 1099551473989 (b) 3127 and (c) 6557 using 3, 4, and 5 qubits, respectively}. We compare the results obtained from experiments (red line) to ideal simulation (blue line), simulation containing phase and amplitude damping noise (orange line), and the residual $ZZ$-coupling (green line). The amplitude damping rate (T1$\approx$64$\mu$s), dephasing rate (T2$\approx$82$\mu$s), and residual $ZZ$-coupling (Table \ref{table:zz-rate-characterization}) reflect the values measured on the quantum processor. We observe that our experimental performance is limited by the residual ZZ-coupling present in the quantum processor.}
    \label{fig:success_rates}
\end{figure*}

\section{Experimental analysis}
\label{sec:experiment}

In this work, we implement the VQF algorithm using the QAOA ansatz on the \emph{ibmq\_boeblingen} superconducting quantum processor (Appendix \ref{appendix:hardware}) to factor three biprime integers: 1099551473989, 3127, and 6557 which are classically preprocessed to instances with 3, 4, and 5 qubits respectively. The number of qubits required for each instance varies based the amount of classical pre-processing performed; 24 iterations of classical pre-processing were used for 1099551473989, 8 iterations for 3127, and 9 iterations for 6557.

In order to optimize the QAOA circuit to find $\boldsymbol{\gamma}_{\mathrm{opt}}$ and $\boldsymbol{\beta}_{\mathrm{opt}}$ we employ a layer-by-layer approach \cite{Zhou2018}. Although there are alternative strategies for training QAOA circuits \cite{Zhou_2020,mcclean2020low}, this approach has been shown to require a grid resolution that scales polynomially with respect to the number of qubits required \cite{Anschuetz2018}. This approach has two phases.

In the first phase, for each layer $k$, we first evaluate a two-dimensional energy landscape by sweeping through different $\gamma_k$ and $\beta_k$ values while fixing the parameters for the previous layers (1, 2, ... through $k-1$) to the optimal values found earlier. In our experiments, for 1099551473989 we evaluate the energy for discrete values of $\gamma_{k}$ and $\beta_{k}$ ranging from $0$ to $2\pi$ with a resolution of $\pi/6$, which yields $144$ circuit evaluations for each layer. For 3127 and 6557, we use a resolution of ~$2\pi/23$, which yields $529$ circuit evaluations for each layer (Appendix \ref{appendix:higher_depth}). We use the degeneracy in the energy landscape caused by evaluating $\beta$ up to $2\pi$ instead of $\pi$ as a feature to better understand the pattern of the energy landscape. We then select the optimal values for $\gamma_k$ and $\beta_k$ from this grid, and combine them with the optimal values from the previous layer. We increment $k$ until it reaches the last layer, in which case the completion of the final round of the above steps marks the end of the first phase of the algorithm. 

In the second phase, we use the values $\{(\gamma_k,\beta_k)\}$ obtained from the first phase as the initial point for a gradient-based optimization using the L-BFGS-B method over all 2$k$ parameters. In order to measure the gradient of $E(\boldsymbol{\gamma}, \boldsymbol{\beta})$ with respect to the individual variational parameters, we use analytical circuit gradients using the parameter-shift rule \cite{Schuld2019}. As shown in Fig.\ \ref{fig:success_rates}, we use the optimal parameters found for a circuit with $p$ layers using this method to prepare and measure $\ket{\boldsymbol{\gamma}_{\rm opt}, \boldsymbol{\beta}_{\rm opt}}$ in order to calculate the algorithm success rate as defined by equation \eqref{eq:success rate}.

The performance of our algorithm relies on the optimization over an energy surface spanned by the variational parameters $\gamma$ and $\beta$ for each layer. Consequently, by studying this energy surface, we can understand the performance of the optimizer in tuning the parameters of the ansatz state, which in turn impacts the success rate. We do so by visualizing the energy landscape for the $k^{\mathrm{th}}$ layer of the ansatz as a function of the two variational parameters, $\beta_k$ and $\gamma_k$, fixing $\boldsymbol{\gamma}=(\gamma_0,..,\gamma_{k-1})$ and $\boldsymbol{\beta}=(\beta_0,..,\beta_{k-1})$ to the optimal values obtained through ideal simulation. Comparing these landscapes between experimental results, ideal simulation, and noisy simulation provides us with a valuable insight into the performance of the algorithm on quantum hardware. 

The main sources of incoherent noise present in the quantum processor are qubit relaxation and decoherence. In order to capture the effects of these sources of noise on our quantum circuit in simulation, we apply a relaxation channel with relaxation parameter $\epsilon_r$ and a dephasing channel with dephasing parameter $\epsilon_d$ after each gate, described by:
\begin{align}\label{eq:relaxation}
   \mathcal{E}(\rho) = \sum_{m=1}^3 E_m \rho E_m^{\dagger}, 
\end{align}
with Kraus operators
\begin{equation} \label{eq:noise_model}
    \begin{split}
     E_1 &= \begin{pmatrix} 
           1 & 0 \\
           0 & \sqrt{1 - \epsilon_r - \epsilon_d }
        \end{pmatrix}, \\ 
    E_2 & = \begin{pmatrix}
            1 & \sqrt{\epsilon_d} \\
            0 & 0
        \end{pmatrix} ,  \\
    E_3 & = \begin{pmatrix}
            1 & 0 \\
            0 & \sqrt{\epsilon_r}
        \end{pmatrix}.
    \end{split}
\end{equation}

Furthermore, a dominant source of coherent error is a residual $ZZ$-coupling between the transmon qubits \cite{Koch2007}. This interaction is caused by the coupling between the higher energy levels of the qubits, and is especially pronounced in transmons due to their weak anharmonicity. In our experiments, we measure the average residual $ZZ$-coupling strength to
be $\xi/2\pi=102 \pm 124$kHz (Appendix \ref{appendix:zz}). While the additional $ZZ$ rotation caused by this interaction when the qubits are idle between gates can be compensated by the tunable parameters $\boldsymbol{\gamma}$, this interaction has a severe impact on the two-qubit gates. In our experiments the Ising terms $e^{i \gamma ZZ}= \mathrm{CNOT} \circ I \otimes Z(\gamma) \circ \mathrm{CNOT}$ are realized using a single-qubit $Z$ rotation conjugated by CNOT gates. Each CNOT is comprised of a $ZX$ term generated by a two-pulse echo cross-resonance gate \cite{PRXQuantum.1.020318} and single-qubit rotations:
\begin{equation}
    \label{eq: cnot implementation}
    \mathrm{CNOT} = e^{-i\frac{\pi}{4}} [ZI]^{-1/2} [ZX]^{1/2} [IX]^{-1/2}
\end{equation}

However, in the presence of residual $ZZ$-coupling, the axis of rotation of the cross-resonance (CR) gate is altered:
\begin{equation}
    \label{eq:zz noise}
    \mathrm{CR}(t) = e^{i \Gamma ZX t} \mapsto e^{i (\Gamma ZX + \Xi) t}
\end{equation}
where $\Gamma$ is the $ZX$ coupling strength between the qubits as the result of the cross-resonance drive and $\Xi$ is the overall effect of the residual ZZ-coupling between the qubits involved in the drive and their neighbors (see Appendix \ref{appendix:zz}). As a result the two-qubit interactions are fundamentally altered, which leads to a source of error that cannot be corrected for through the variational parameters.

Fig. \ref{fig:energy landscapes - zz} shows the energy landscapes from factoring 1099551473989 across 3 layers. Comparing the the energy landscape produced in experiment with the ideal simulation and noisy simulations indicates that the energy landscapes produced in experiment is in close agreement with those produced by the ideal simulation and the simulation with incoherent noise for $p=1$ and $p=2$. However, for $p=3$ the energy landscapes produced in experiment have considerable deviations from the ideal simulations due to coherent errors. The presence of incoherent sources of noise do not change the energy landscape pattern, and only reduce the contrast at every layer. On the other hand, the residual $ZZ$-coupling between qubits alters the pattern of the energy landscape. While at lower depths the impact of this noise source is minor, yet as we add more layers of the ansatz the effects accumulate, constructively interfere, and amplify due to the coherent nature of the error. 

\begin{figure}[b]
    \centering
    \includegraphics[width=8cm]{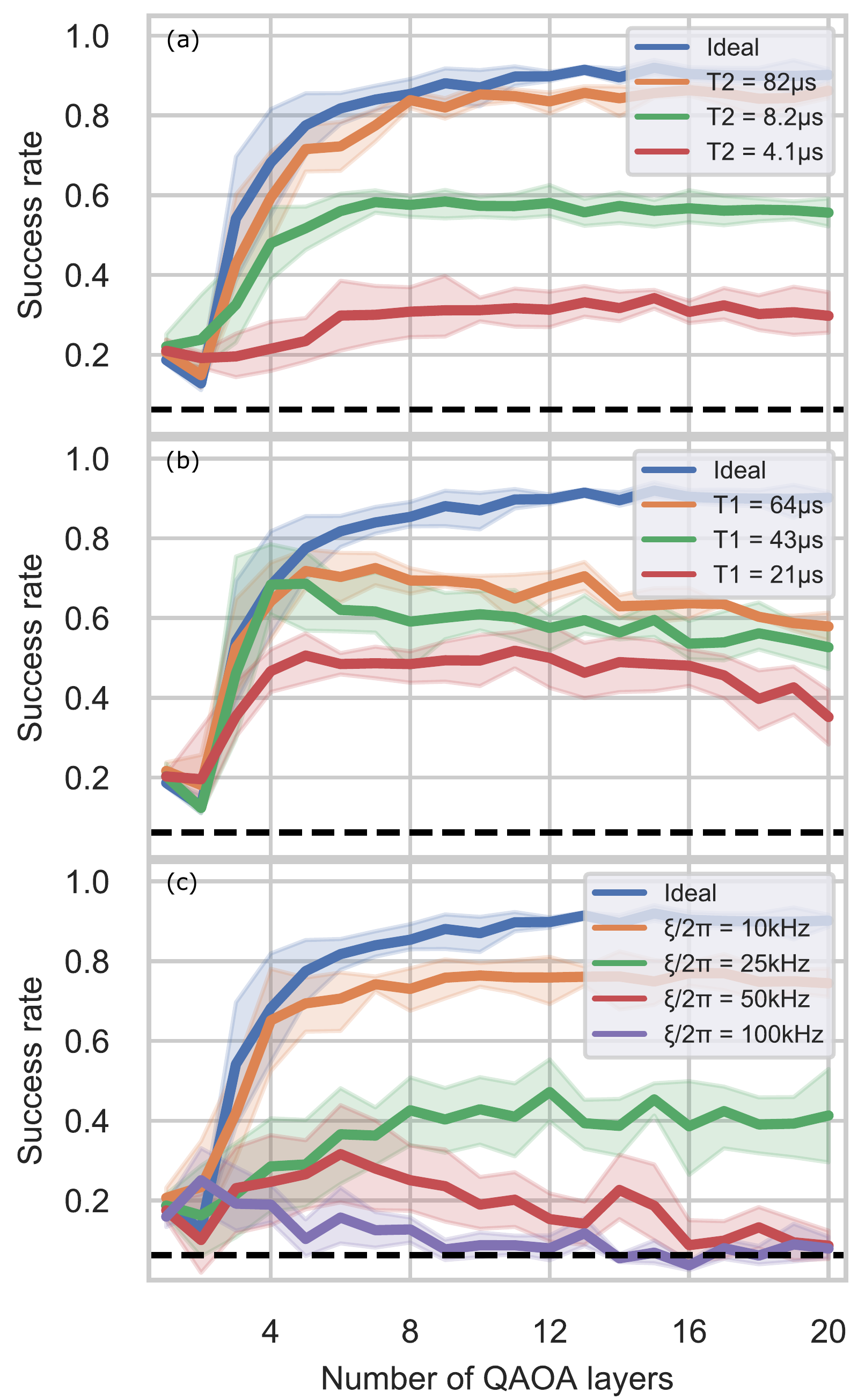}
    \caption{\textbf{Impact of different sources of noise on success rate when factoring 6557 using 5 qubits: }(a) phase damping, (b) amplitude damping, and (c) residual $ZZ$-coupling between qubits with a two-qubit gate time of $t_g=315$ ns. The average T2, T1, and frequency of the residual $ZZ$-coupling for the given device are $82 \mu$s, $64 \mu$s, and $102$ kHz respectively.}
    \label{fig:all_noise_sources}
\end{figure}

Fig. \ref{fig:success_rates} reports the average success rate of VQF as a function of the number of layers in the ansatz for experiment and simulation. 
As would be expected, in the ideal case the average success rate approaches the ideal value of 1 with increasing layers of the ansatz for each of the instances studied. Comparing experimental results to those obtained from ideal simulation shows a large discrepancy in the success rate. This discrepancy is not explained by the incoherent noise caused by qubit relaxation and dephasing; the coherent error caused by the residual $ZZ$-coupling is the dominant effect negatively impacting the success rate. Therefore we conclude that the performance of VQF, and other QAOA-based hybrid quantum-classical algorithms by extension, is limited by the residual $ZZ$-coupling between the qubits on transmon-based superconducting quantum processors.

We investigate the impact of different noise channels on the success rate scaling of VQF, with results shown in Fig. \ref{fig:all_noise_sources}. In the ideal scenario, with every added layer of QAOA, which increases the ansatz expressibility \cite{Sim2019}, the algorithm success rate increases. However, in the presence of phase damping, after an initial increase in the success rate we observe a plateau after which the succcess rate neither increases nor decreases as more layers of the ansatz circuit are added. We expect that this behavior is caused by a loss of quantum coherence at higher layers of circuit, which hinders the quantum interference required for the algorithm. In constrast, while undergoing amplitude damping, after an initial increase in the success rate we observe a steady decay in the algorithm's performance. This decay is a result of qubit relaxation back to the $\ket{0}$ state. When examining the effect of the coherent error induced by the residual $ZZ$-coupling, we see a more dramatic effect than that induced by either phase damping or amplitude damping. In the presence of the coherent error induced by the residual $ZZ$-coupling, we observe a plateau in the success rate for lower frequencies (10 kHz and 25 kHz) and a decline in success rate for higher frequencies (50 kHz and 100 kHz) within the first twenty layers of the ansatz.

While each of the sources of noise impact the algorithm's performance, the residual $ZZ$-coupling between the qubits in the quantum processor has the most dominant impact on the success rate; Our analysis indicates that this coherent error is the main source of error impeding the effectiveness of the algorithm. The accumulation of coherent noise from this source significantly alters the ansatz at higher layers, and cannot be corrected using the variational parameters. The results shown in Fig.\ \ref{fig:all_noise_sources} indicate that the performance of VQF can be significantly improved by engineering $ZZ$ suppression \cite{Sung2020, Kandala2020}.
Since VQF is a non-trivial instance of QAOA we expect this performance analysis to also hold for other quantum algorithms based upon QAOA.

\section{Conclusion}
\label{sec:conclusion}

In this work, we analyzed the performance of VQF on a fixed-frequency transmon superconducting quantum processor and investigated several kinds of classical and quantum resource tradeoffs. We map the problem of factoring to an optimization problem and use variable amounts of classical preprocessing to adjust the number of qubits required. We then use the QAOA ansatz with a variable number of layers to find the solution to the optimization problem. We find that the success rate of the algorithm saturates as $p$ increases, instead of decreasing to the accuracy of random guessing. While more layers increases the expressibility of the ansatz, at higher depths the circuit suffer from the impacts of noise.

Our analysis indicate that the residual $ZZ$-coupling between the transmon qubits significantly impacts performance. While relaxation and decoherence have an impact on the quantum coherence, especially at the deeper layers of QAOA, the effect of coherent noise can quickly accumulate and limit performance. By developing a noise model incorporating the coherent sources of noise, we are able to much more accurately predict our experimental results.

 There have been various techniques \cite{Proctor2020,Cross2019} proposed for benchmarking the capability of a quantum device for performing arbitrary unitary operations. However, recent findings on experimental systems \cite{Kjaergaard} suggest that existing benchmarking methods may not be effective for predicting the capability of a quantum device for specific applications. This has motivated recent works that focus on benchmarking the performance of a quantum device for applications such as generative modeling \cite{Benedetti2019} and fermionic simulation \cite{2003.01862}. Our study of VQF on a superconducting quantum processor has the potential to become another entry in the collection of such ``application-based" benchmarks for quantum computers. 

\bibliography{refs}

\section*{Acknowledgments}
The authors would like to thank Eric Anschuetz, Morten Kjaergaard, Sukin Sim, Peter Johnson, and Jonathan Olson for fruitful discussion. We thank Maria Genckel for helping with the graphics. We would also like to thank the IBM Quantum team for providing access to the \emph{ibmq\_boeblingen} system, and for help debugging jobs. We acknowledge the use of IBM Quantum services for this work. The views expressed are those of the authors, and do not reflect the official policy or position of IBM.

\section*{Author Contribution}
Y.\ C.\ conceived the project, 
A.\ K.\ and W.\ S.\ developed and executed the quantum workflow for measurements and simulations, 
A.\ K.\ and B.\ P.\ developed the noise model, 
T.\ S.\ assisted with Qiskit and IBM Q quantum backend support as relates to VQF.
All authors contributed to writing the manuscript.

\section*{Data Availability}
The raw data used for generating the plots in this paper are available upon request.

\section*{Competing Interests}
The authors declare no competing interests for the work in this paper.

\clearpage
\onecolumngrid

\appendix

\section{Device Information}\label{appendix:hardware}

\begin{figure}[h!]
    \centering
    \includegraphics[width=8cm]{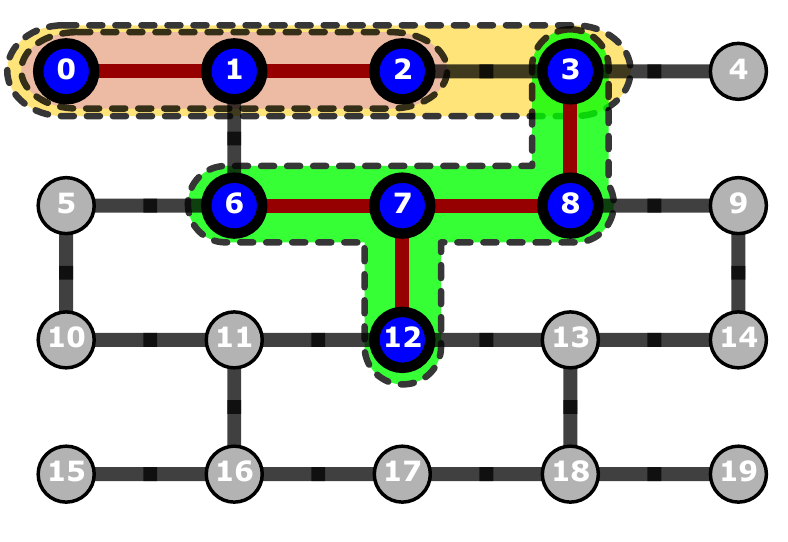}
    \caption{\textbf{IBMQ Boeblingen connectivity map}. To factor 1099551473989 we use qubits 0, 1, and 2 (shaded in maroon). To factor 3127 we use qubits 0, 1, 2, and 3 (shaded in yellow). To factor 6557 we use qubits 3, 6, 7, 8, and 12 (shaded in green).} 
    \label{fig:boeb_map}
\end{figure}

For our experiments we use the 20 qubit Boeblingen system. This quantum processor consists of 20 fixed-frequency transmon qubits, with microwave-driven single and two-qubit gates, and with a connectivity map as depicted in Fig. \ref{fig:boeb_map}. For the experiments discussed in this paper use the CNOT as our native two-qubit gate which is driven via the cross-resonance interaction between coupled qubits. Experiments were conducted on the highlighted qubits, for their respective single and two qubit gate fidelities have a higher fidelity, as well as a relatively low readout error. Qubit properties and two-qubit gate characterization can be found in table I and II respectively. For factoring 1099551473989 we use Q0, Q1, and Q2, for factoring 3127 we use Q0, Q1, Q2, and Q3, and for factoring 6557 we use Q3, Q6, Q7, Q8, and Q12.

\begin{table}[h!]
\centering
\label{table:qubit-characterization}
\begin{tabular}{| c | l| l | l | l |l | l | l | l | l | l | l | l | }
\toprule
  \textbf{Property} & &          Q0  &          Q1  &          Q2  &          Q3  &          Q4  &          Q5  &          Q6  &          Q7  &          Q8  &          Q9  &          Q12 \\ \hline 
\hline
1QB Gate Error & mean &  4.28e-04 &  3.08e-04 &  2.83e-04 &  4.03e-04 &  7.07e-04 &  5.19e-04 &  7.37e-04 &  3.16e-04 &  4.20e-04 &  4.08e-04 &  3.47e-04 \\
   & std &  1.56e-04 &  5.09e-05 &  7.98e-05 &  6.61e-05 &  4.13e-04 &  1.05e-04 &  5.230e-04 &  4.80e-05 &  8.72e-05 &  6.20e-05 &  3.54e-05 \\ \hline 
Frequency (GHz) & mean &  5.05 &  4.85 &  4.70 &  4.77 &  4.37 &  4.91 &  4.73 &  4.55 &  4.66 &  4.77 &  4.74e \\
   & std &  2.42e-06 &  4.35e-06 &  3.18e-06 &  3.73e-06 &  1.01e-05 &  9.48e-05 &  1.49e-05 &  4.26e-06 &  4.84e-06 &  6.37e-05 &  3.26e-06 \\ \hline 
Readout Error (\%)& mean &  2.48 &  2.73 &  3.68 &  2.28 &  5.13 &  1.95 &  4.69 &  3.80 &  5.76 &  6.34 &  4.59 \\
   & std &  .523 &  .348 &  .964 &  .373 &  4.27 &  .711 &  1.54 &  .805 &  .352 &  2.43 &  .465 \\ \hline
T1 ($\mu$s)& mean &  62.8 &  59.1 &  105 &  80.2 &  87.3 &  80.0 &  64.2 &  81.0 &  44.8 &  57.5 &  80.5 \\
   & std &  21.3 &  11.5 &  21.7 &  8.98 &  17.4 &  19.5 &  15.0 &  14.0 &  12.4 &  5.61 &  17.7 \\ \hline 
T2 ($\mu$s) & mean &  97.4 &  86.5 &  104 &  42.7 &  76.9 &  54.2 &  75.0 &  88.9 &  69.6 &  84.4 &  107 \\
   & std &  38.6 &  21.9 &  31.2 &  4.38 &  32.3 &  16.3 &  20.7 &  17.7 &  18.1 &  40.0 &  26.1 \\ \hline
\end{tabular}
\caption{\textbf{Single-qubit characterization}. Data acquired over the span of 14 days.}
\end{table}

\begin{table}[h!]
\centering
\label{table:two-qubit-gates}
\begin{tabular}{ |l|c|c|} 
\toprule
Coupling & Error & Time (ns)    \\ 
 \hline
 \hline
 Q0-Q1 \: & \: 7.38e-03 $\pm$ 9.7e-04 &220 \\ 
 \hline
 Q1-Q2 \: & \: 6.87e-03 $\pm$ 9.9e-04 \: & 334 \\ 
 \hline
 Q2-Q3 \: & \: 9.75e-03 $\pm$ 20.8e-04 \: & 277 \\ 
 \hline
 Q6-Q7 \: & \: 13.55e-03 $\pm$ 55.8e-04 \: & 256 \\ 
 \hline
 Q7-Q8 \: & \: 10.76e-03 $\pm$ 10.4e-04 \: & 412 \\
 \hline 
 Q7-Q12 \: & \: 14.96e-03 $\pm$ 86.1e-04 \: & 306 \\  
 \hline
 Q8-Q3 \: & \: 10.06e-03 $\pm$ 12.3e-04 \: & 363 \\  
 \hline
\end{tabular}
\caption{\textbf{Two-qubit properties}. Data acquired over the span of 14 days.}
\end{table}

\section{Residual ZZ-coupling}\label{appendix:zz}
In addition to the well-known incoherent processes of relaxation and dephasing, there are coherent noise mechanisms that may affect the overall algorithm performance. Taking into account these coherent errors has become increasingly important in near term algorithms, for their effects cannot be captured by standard benchmark techniques (such as randomized benchmarking), which may impact the overall performance of a quantum algorithm. For superconducting qubits in general, and especially for fixed-frequency transmons, these coherent sources of error are qubit crosstalk due to microwave leaking out from one qubit to another while driving single qubit gates, and residual ZZ-coupling between connected qubits, due to the relatively small anharmonicity of the transmons. Qubit crosstalk is relatively harmless in variational algorithms, for its effect, namely single qubit overrotations, can be learned and cancelled out by the classical optimizer on each optimization step. However, the residual ZZ noise which accumulates during the gates cannot be learned by the optimizer, as it doesn't commute with the native ZX term from the CR gate, and needs to be taken into account in the quantum circuit. 

\begin{figure}[t]
    \centering
    \includegraphics[width=6cm]{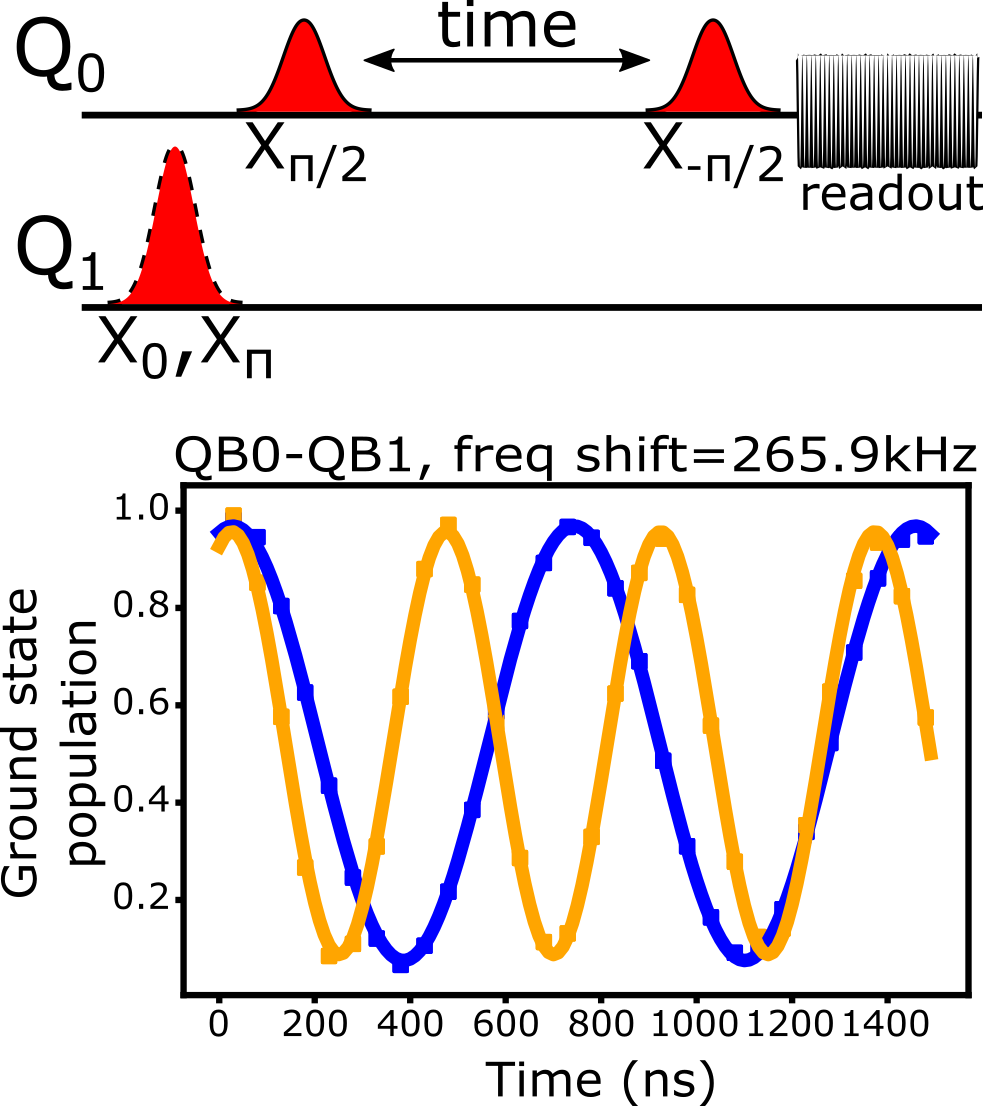}
    \hspace{.5em}
    \centering
    \caption{\textbf{Measuring the strength of the residual $ZZ$-coupling between pairs of qubits.} For each of the pairs of qubits, we measure the Ramsey oscillation frequency of each qubit when the coupled qubit in question is in the $\ket{0}$ state and in the $\ket{1}$ state. The strength of the residual $ZZ$-coupling is then the difference between these frequencies.}
    \label{fig: ZZ-rate}
 \end{figure}

\begin{table}[t]
\centering
\label{table:zz-rate-characterization}
    \begin{tabular}[b]{ |l|c|} 
    \toprule
      Coupling \: & \: $\xi/2\pi$ (kHz) \\\hline \hline
      Q0-Q1 \: & \: 275.1 $\pm$  6.7 \\  \hline
      Q1-Q2 \: & \: 226.7 $\pm$  3.4 \\  \hline
      Q1-Q6 \: & \: 66.1 $\pm$  5.8 \\  \hline
      Q2-Q3 \: & \: 86.3 $\pm$  0.8 \\  \hline
      Q3-Q4 \: & \: -116.0 $\pm$  3.5 \\  \hline
      Q3-Q8 \: & \: 54.7 $\pm$  1.5 \\  \hline
      Q5-Q6 \: & \: 186.4 $\pm$  3.7 \\  \hline
      Q5-Q10 \: & \: 81.1 $\pm$  1.6 \\  \hline
      Q6-Q7 \: & \: 117.8 $\pm$  5.2 \\  \hline
      Q7-Q8 \: & \: 69.8 $\pm$  0.2 \\  \hline
      Q7-Q12 \: & \: 74.9 $\pm$  13.2 \\  \hline
    \end{tabular}
    \begin{tabular}[b]{ |l|c|} 
    \toprule
      Coupling \: & \: $\xi/2\pi$ (kHz) \\\hline \hline
      Q8-Q9 \: & \: 135.6 $\pm$  0.6 \\  \hline
      Q10-Q11 \: & \: -249.8 $\pm$  0.4 \\  \hline
      Q11-Q12 \: & \: 242.6 $\pm$  3.7 \\  \hline
      Q11-Q16 \: & \: 79.9 $\pm$  0.4 \\  \hline
      Q12-Q13 \: & \: 79.3 $\pm$  3.0 \\  \hline
      Q13-Q14 \: & \: 67.7 $\pm$  3.3 \\  \hline
      Q13-Q18 \: & \: 53.9 $\pm$  2.9 \\  \hline
      Q15-Q16 \: & \: 119.5 $\pm$  3.2 \\  \hline
      Q16-Q17 \: & \: 108.0 $\pm$  1.6 \\  \hline
      Q17-Q18 \: & \: 98.8 $\pm$  13.4 \\  \hline
      Q18-Q19 \: & \: 383.6 $\pm$  13.3 \\  \hline
    \end{tabular}
\caption{\textbf{Reporting the measured strength of the residual $ZZ$-coupling between all pairs of qubits.} Data is averaged over two trials per connected pair of qubits. }
\end{table}

In order to estimate the effect of the residual ZZ-coupling in actual experiments, it is crucial to derive an accurate Hamiltonian model that captures the effect of higher energy levels in the energy spectrum. To this end, we sketch the derivation of an effective Hamiltonian for two transmon qubits interacting through a common transmission line resonator (which corresponds to qubits connected through solid lines in Fig. \ref{fig:boeb_map}). We model the transmon qubit $j$ as a Duffing oscillator
\begin{equation}
H_{\text{qb}_i}= \omega_0(b_i^\dagger b_i+1/2)+\frac{\delta_i}{2}b_i^\dagger b_i(b_i^\dagger b_i-1),
\end{equation}
that interacts with other transmons via the transmission line resonator through the exchange of virtual excitations
\begin{equation}
    H = \omega_{\text{res}}a^\dagger a+H_{\text{qb}_{1}}+ H_{\text{qb}_{2}}+ \sum_{j=1}^2 g_j(a^\dagger b_j+H.c),
\end{equation}
where $g_j$ represents the coupling constant between transmon $j$ and the resonator. We can diagonalize this Hamiltonian through a Schrieffer-Wolff transformation to adiabatically remove the resonator effect, getting an effective transmon-transmon interaction that reads 
\begin{equation}
    H = \sum_{j=1}^2 \omega_j b_j^\dagger b_j + J(b_1^\dagger b_2+b_2^\dagger b_1),
\end{equation}
where $J=-g_1g_2\times(\Delta_1+\Delta_2)/2(\Delta_1\Delta_2)$, is the effective coupling between transmons, and  $\Delta_j=\omega_{\text{res}}-\omega_{j}$ the detuning between the transmon $j$ and resonator.
We now bring the above Hamiltonian to its diagonal form, as it represents
the physical basis where the qubits are actually measured, and project it to the two-qubit subspace, yielding
\begin{equation}
    H = \sum_{j=1}^2 \tilde{\omega}_j Z_j + \xi Z_1 Z_2,
\end{equation}
where $\tilde{\omega}_j$ are (dressed) qubit frequencies that are actually measured in an experiment, and $\xi = 2J^2(\delta_1+\delta_2/(\delta_1-\Delta)(\delta_2+\Delta))$. The residual ZZ-coupling between two qubits, $Q_i$ and $Q_j$, can be experimentally measured by finding the difference in the Ramsey oscillation frequency of $Q_i$ while $Q_j$ is in the $\ket{0}$ and $\ket{1}$ state (Fig. \ref{fig: ZZ-rate}). The residual ZZ-coupling between the qubits used in our experiments can be found in Table \ref{table:zz-rate-characterization}.

In order to model the effects of the residual ZZ-coupling in simulation, we implement the ECR$_{2-pulse}$ protocol as follows:
\begin{equation}
  \label{eq:ecr 2-pulse}
    \mathrm{ECR_{2-pulse}(t) = [XI] CR(-t/2) [XI] CR(t/2)}
\end{equation}

As mentioned, in the presence of the residual ZZ-coupling between qubits, the axis of rotation for the cross-resonance gate is fundamentally altered. When accounting only for the residual ZZ-coupling coupling between the qubits involved in the cross-resonance gate, we model the effect as follows:
\begin{equation}
  \label{eq: overall zz effect without spectators}
    \mathrm{\Xi = \xi_{c,t} ZZ}
\end{equation}
where $\xi_{c,t}$ is the strength of the residual ZZ-coupling between the control qubit and the target qubit. 

However, the presence of residual ZZ-couplings between each of these two qubits and its respective spectator qubits also alters the effective cross-resonance gate. When accounting for the involved spectator qubits, we model the effect as:
\begin{equation}
  \label{eq: overall zz effect with spectators}
    \mathrm{\Xi = \xi_{c,t} ZZ_{c,t} + \sum_{i=0}^U \xi_{c,i} ZZ_{c,i} + \sum_{j=0}^V \xi_{t,j} ZZ_{t,j}}
\end{equation}
where U and V are the set of spectator qubits of the control qubit and target qubit respectively. All values of $\xi$ are as reported in Table \ref{table:zz-rate-characterization}. 

\begin{figure}[h!]
    \centering
    \includegraphics[height=2.5in]{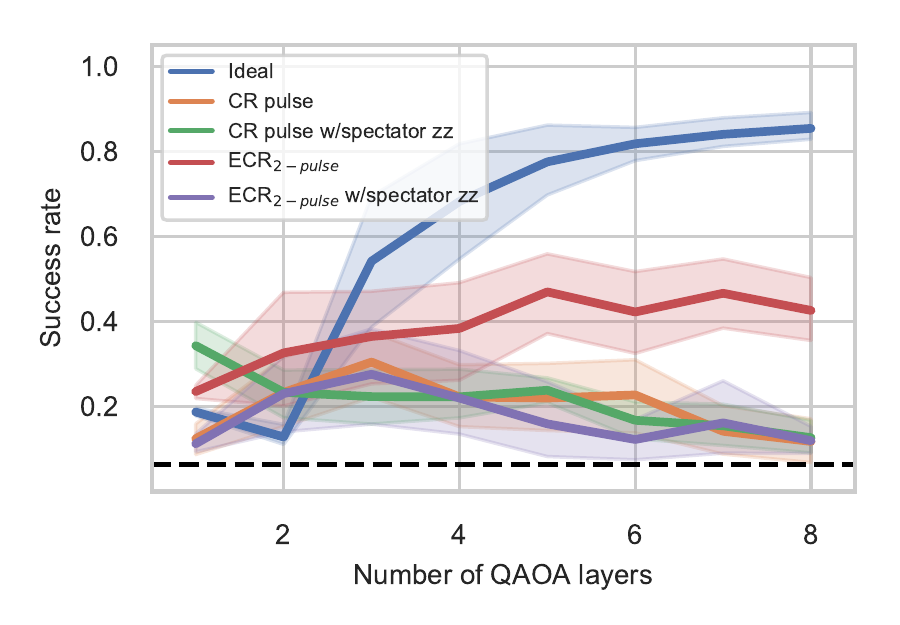}
    \caption{\textbf{Effects of residual ZZ-coupling on spectator qubits with different cross-resonance gate decompositions}. We compare the effects of incorporating the residual ZZ-coupling noise on spectator qubits during the cross resonance gate under both the single CR pulse and ECR$_{2-pulse}$ protocols.}
    \label{fig: different effects of residual zz-coupling modeling}
\end{figure}

In Fig. \ref{fig: different effects of residual zz-coupling modeling} we show how different implementations of the effect of the residual ZZ-couplings impact the success rate of VQF on 6557. We note that when spectator qubits are not considered, the ECR$_{2-pulse}$ protocol shows significant improvement over the single cross-resonance gate. However, when the residual ZZ-coupling of spectator qubits is considered, this improvement is lost and the success rates trend rapidly toward the random success rate. 

\section{VQF Preprocessing }\label{appendix:preprocessing}

By simplifying the clauses using binary rules we reduce the quantum resources required for executing the VQF algorithm. The classical preprocessing procedure iterates through all clauses $\{C_i\}$ a constant number of times, using a set of binary rules to solve or make deductions where possible. Assuming $x,y,z \in F_2, a \in Z^+$  we apply the following classical preprocessing rules:

\begin{enumerate}
    \item $xy=1 \Rightarrow x=y=1$
    \item $x+y=1 \Rightarrow xy=0$
    \item $x+y=2z \Rightarrow x=y=z$
    \item $\sum_{i=1}^a x_i = a \Rightarrow x_i=1$
    \item if $x_i=1$ violates $\text{max}(lhs)=\text{max}(rhs) \Rightarrow x_i=0$
    \item if $\text{min}(lhs)<\text{min}(rhs)$ and $x$ is the only variable on lhs $\Rightarrow x=1$
    \item the parity of the $lhs$ must be the same as the $rhs$
\end{enumerate}

\begin{figure}
    \centering
    \includegraphics[width=8cm]{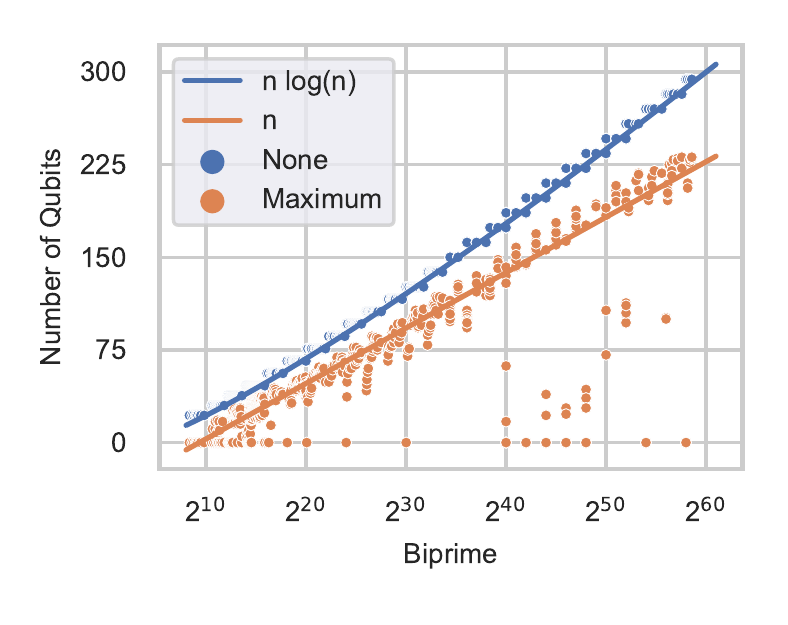}
    \caption{\textbf{Empirical results on number of qubits required for factoring a biprime $N$ after classical preprocessing.} $n$ represents the number of bits needed to represent the given biprime.}
    \label{fig:pre-processor}
\end{figure}

\begin{figure}
    \centering
    \includegraphics[width=8cm]{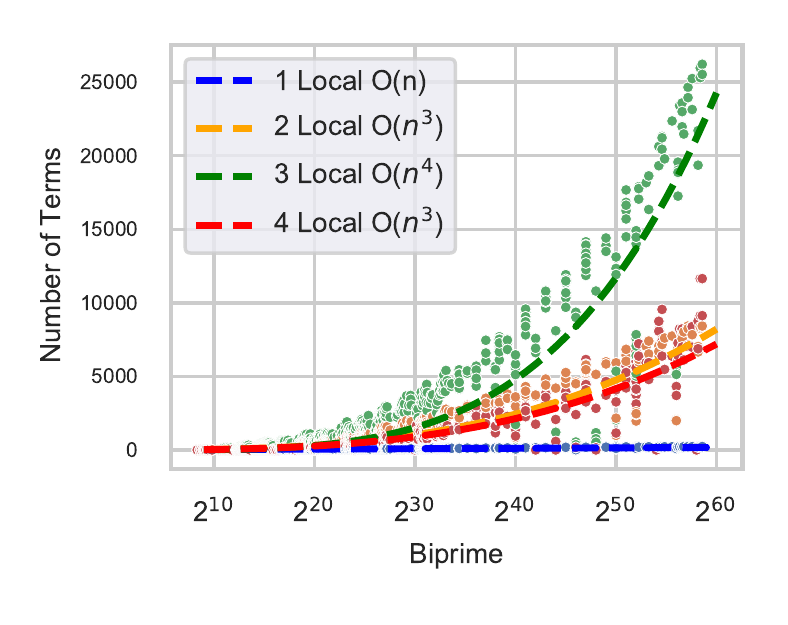}
    \caption{\textbf{Empirical results on number of n-local terms in the hamiltonain for factoring a biprime $N$ after classical preprocessing.}}
    \label{fig:n-local_term}
\end{figure}

The runtime of the classical preprocessor is $O(n^2)$ \cite{Anschuetz2018}. Without preprocessing the VQF qubit requirements scale as $O(n$log$\:n)$, whereas this resource bound empirically gets reduced to $O(n)$ using the classical preprocessor (Fig. \ref{fig:pre-processor}). We additionally note that for certain biprimes, the preprocessor is able to significantly reduce the number of required qubits, in some cases completely solving the clauses. Additionally, in Fig. \ref{fig:n-local_term}, we show the scaling of 1-local, 2-local, 3-local, and 4-local terms in the resulting hamiltonians, after classical preprocessing, for factoring a given biprime $N$. 

\section{Circuit Metrics}

\begin{table}[h!]
\centering
\begin{tabular}{ |l|c|c|c|} 
\toprule
 & CNOT gates & Single-qubit gates & Circuit depth \\ 
 \hline
 \hline
 1st layer & 4 & 22 & 13 \\ 
 \hline
 2nd layer & 8 & 41 & 25 \\ 
 \hline
 3rd layer & 12 & 60 & 37 \\  
 \hline
 4th layer & 16 & 79 & 49 \\ 
 \hline 
 5th layer & 20 & 98 & 61 \\  
 \hline
 6th layer & 24 & 117 & 73 \\  
 \hline
 7th layer & 28 & 136 & 85 \\  
 \hline
 8th layer & 32 & 155 & 97 \\  
 \hline
\end{tabular}
\caption{\textbf{Properties of QAOA circuits 1099551473989 (3 qubits)}. The qubit mapping used for this instance is: [0, 1, 2].}
\label{table:3-qubit-metrics}
\end{table}

\begin{table}[h!]
\centering
\begin{tabular}{ |l|c|c|c|} 
\toprule
 & CNOT gates & Single-qubit gates & Circuit depth \\ 
 \hline
 \hline
1th layer & 18 & 12 & 21 \\ 
 \hline
2th layer & 37 & 20 & 39 \\ 
 \hline
3th layer & 59 & 28 & 58 \\ 
 \hline
4th layer & 81 & 36 & 77 \\ 
 \hline
5th layer & 103 & 44 & 96 \\ 
 \hline
6th layer & 125 & 52 & 115 \\ 
 \hline
7th layer & 147 & 60 & 134 \\ 
 \hline
8th layer & 169 & 68 & 153 \\ 
 \hline
\end{tabular}
\caption{\textbf{Properties of QAOA circuits 3127 (4 qubits)}. The qubit mapping used for this instance is: [0, 1, 2, 3].}
\label{table:4-qubit-metrics}
\end{table}

\begin{table}[h!]
\centering
\begin{tabular}{ |l|c|c|c|} 
\toprule
 & CNOT gates & Single-qubit gates & Circuit depth \\ 
 \hline
 \hline
 1st layer & 13 & 44 & 25 \\ 
 \hline
 2nd layer & 29 & 86 & 52 \\ 
 \hline
 3rd layer & 45 & 128 & 76 \\  
 \hline
 4th layer & 94 & 203 & 116 \\ 
 \hline 
 5th layer & 95 & 230 & 132 \\  
 \hline
 6th layer & 147 & 308 & 176 \\  
 \hline
 7th layer & 147 & 334 & 197 \\  
 \hline
 8th layer & 162 & 375 & 226 \\  
 \hline
\end{tabular}
\caption{\textbf{Properties of QAOA circuits on 6557 (5 qubits)}. The qubit mapping used for this instance is: [6, 7, 12, 8, 3].}
\label{table:5-qubit-metrics}
\end{table}

In Tables \ref{table:3-qubit-metrics}, \ref{table:4-qubit-metrics}, and \ref{table:5-qubit-metrics}, we show the relevant properties for the circuits used in both simulation and experiment for each of the problems studied. We use the Qiskit transpiler, with the optimization level set to 1, to create the circuits with the correct gate set for IBMQ devices. Due to the non-deterministic nature of this transpiler, we independently run the transpilation process 10 times, using the circuit with the least number of cx gates.

\section{297491 (4 qubits)}
\label{appendix:297491}

\begin{figure}[h!]
    \centering
    \includegraphics[height=2in]{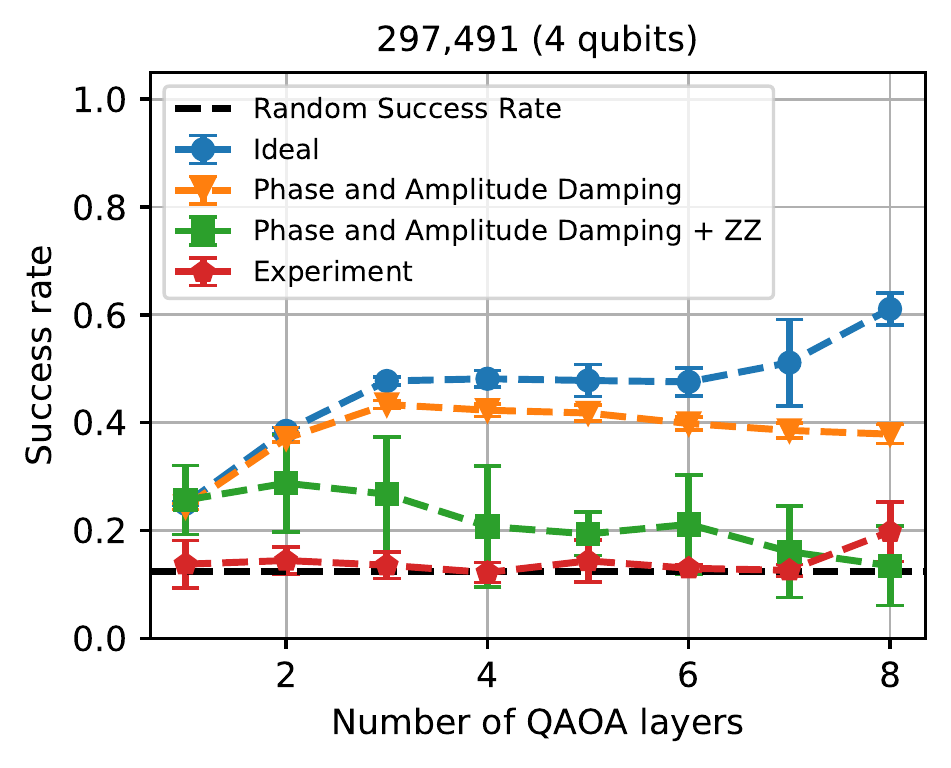}
    \caption{\textbf{Success rates for running VQF on the integer 297491 using 4 qubits}. The qubits used for this instance are 6, 7, 8, and 12. We compare the results obtained from experiments (red line) to ideal simulation (blue line), simulation containing phase and amplitude damping noise (orange line), and the residual $ZZ$-coupling (green line). The amplitude damping rate (T1$\approx$64$\mu$s), dephasing rate (T2$\approx$82$\mu$s), and residual $ZZ$-coupling (Table \ref{table:zz-rate-characterization}) reflect the values measured on the quantum processor.}
    \label{fig:297491}
\end{figure}

In Fig. \ref{fig:297491}, we show the results of running the VQF algorithm on the integer 297491 using 4 qubits (6, 7, 8, and 12). Similar to the results shown in Fig. \ref{fig:success_rates}, we present the success rate of the VQF algorithm as a function of the number of layers in our QAOA ansatz and compare results from experiment with those from simulation with different noise models. In contrast to results shown in Fig. \ref{fig:success_rates}, we observe no improvement in the success rate during experiment versus random sampling. Additionally, the residual $ZZ$-coupling between qubits does not alone explain the degradation in the success of the algorithm when run on the quantum processor. This suggests the presence of an additional source of noise that is unaccounted for, even though the qubits used in this instance are a subset of the ones used to factor 6557.

\section{Higher depth energy landscapes}
\label{appendix:higher_depth}

\begin{figure*}[t]
    \centering
    \includegraphics[width=16cm]{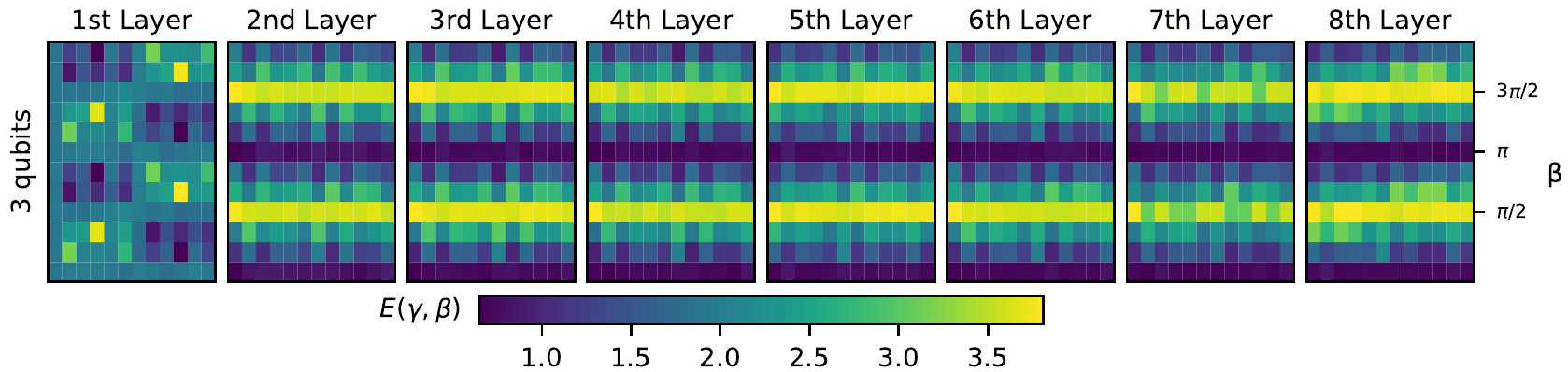}
    \includegraphics[width=16cm]{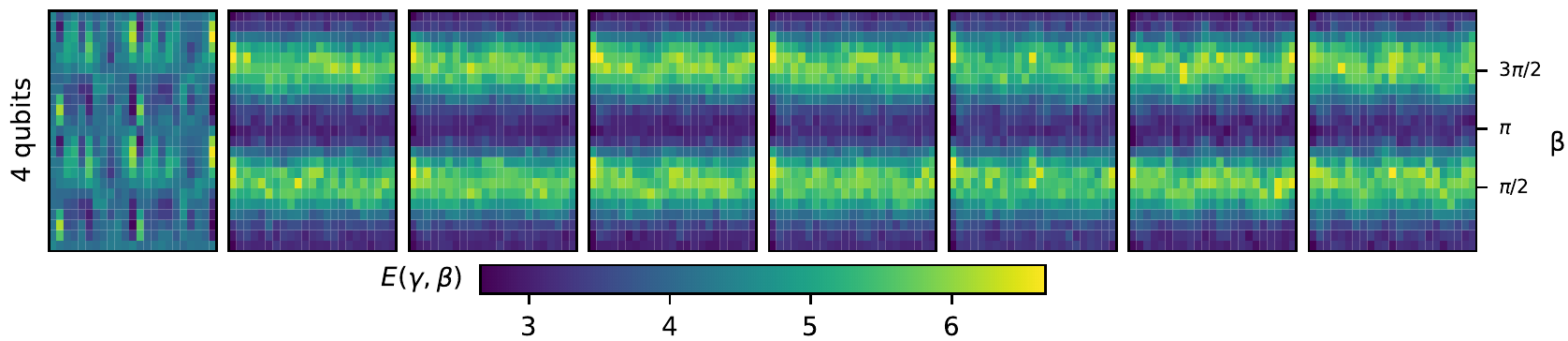}
    \includegraphics[width=16cm]{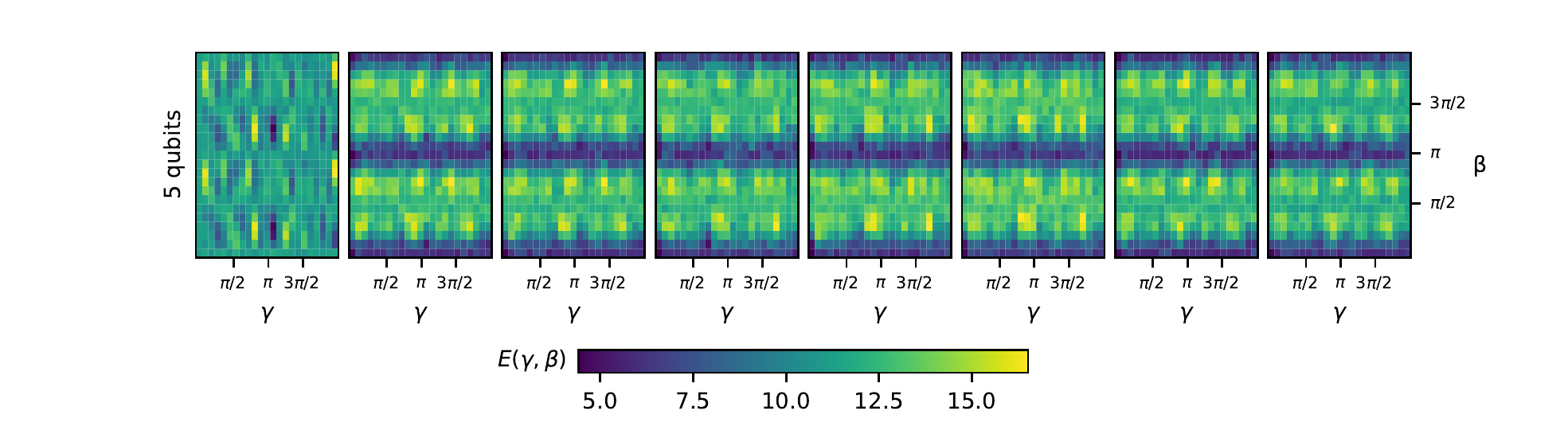}
    \caption{\textbf{QAOA optimization landscapes}. For each VQF instance (rows), we show the resulting optimization landscape as a function of the number of QAOA layers in the ansatz (columns). For the 3 qubit instance (upper), the landscapes are obtained with a 12x12 grid, resulting in 144 uniformly spaced points. Whereas the landscapes obtained for the 4 and 5 qubit instances (middle and bottom, respectively) are created using a 23x23 grid with 529 uniformly spaced points.} 
    \label{fig:high depth energy landscapes}
\end{figure*}

In Fig. \ref{fig:high depth energy landscapes} we display the energy landscapes for the instances representing 1099551473989, 3127, and 6557 using 3, 4, and 5 qubits respectively for $p=(1,2,\cdots,8)$. We observe that the landscapes for $p=(2,\cdots,8)$ follow a similar pattern, not only between layers, but across instances as well. Additionally, we note that we continue to observe structure in the landscapes for high-depth circuits. As reported in Table \ref{table:5-qubit-metrics}, the landscape produced for 6557 with 8 layers requires approximately 162 CNOT gates, 375 single-qubit gates, and a circuit depth of 226.

\end{document}